\documentstyle[pre,manuscript,aps]{revtex} 

\begin{document}
\draft

\title{Strong electrostatic interactions in spherical colloidal systems}

\author{Ren\'{e} Messina\thanks{ email: messina@mpip-mainz.mpg.de }, Christian
  Holm\thanks{ email: holm@mpip-mainz.mpg.de }, and Kurt Kremer\thanks{ email:
    k.kremer@mpip-mainz.mpg.de }}

\address{Max-Planck-Institut f\"{u}r Polymerforschung, Ackermannweg 10, 55128,
  Mainz, Germany}

\date{\today{}}

\maketitle
\begin{abstract}
  We investigate spherical macroions in the strong Coulomb coupling regime
  within the primitive model in salt-free environment.  We first show that the
  ground state of an isolated colloid is naturally overcharged by simple
  electrostatic arguments illustrated by the Gillespie rule. We furthermore
  demonstrate that in the strong Coulomb coupling this mechanism leads to
  ionized states and thus to long range attractions between like-charged
  spheres. We use molecular dynamics simulations to study in
  detail the counterion distribution for one and two highly charged colloids
  for the ground state as well as for finite temperatures.  We compare our
  results in terms of a simple version of a Wigner crystal theory and find
  excellent qualitative and quantitative agreement.\\

\end{abstract}


\narrowtext

\section{Introduction}

Charged colloidal suspensions are often encountered in the everyday life (technology,
biology, medicine ...) and have an important practical impact \cite{Everett_book_1988}.
In numerous application-oriented situations, electrostatic repulsion among colloids
(macroions) is desired in order to obtain a stabilized suspension. Consequently
the understanding of the electrostatic interaction in such systems is motivated
by practical as well as theoretical interests. There is recent experimental
evidence that the effective interaction between two like-charged spherical colloids
(in the presence of neutralizing salts) can be attractive in the presence of
one or two glass walls \cite{Kepler_al_PRL_1994,Crocker_al_PRL_1996,Larsen_al_Nature_1997}.
This is in contrast with the classical work of Derjaguin, Landau, Verwey and
Overbeek (DLVO) based on a linearized Poisson Boltzmann theory \cite{DL_1941,VO_1948},
which foresees only repulsive effective Coulomb forces between two like-charged
spheres even in confined geometry. There are some indications that this attraction
might be explainable in terms of hydrodynamic effects induced by the walls \cite{Squires_PRL_2000}.

Already in the bulk case there have been disputes for a long time about the
existence of long range attractive forces, triggered mainly by the observation
of voids in colloidal solutions\cite{Ise_JCP_1982,Ito_Science_1994,Tata_PRL_1997,Roij_PRL_1997}.
There is no clear experimental and theoretical picture, either, and there have
been speculations that the experiments observed phase coexistence. Recent theoretical
\cite{Shklowskii_PRL_1999a,Netz_EPL_1999,Tokuyama__PRE_1999} and simulation
\cite{Jensen_Physica_1998,Elshad_PRL_1998,Wu_JCP_1999,Linse_PRL_1999} investigations
have shown the existence of \textit{short range} attraction.

In two short communications\cite{Messina_PRL_2000,Messina_EPL_2000}, we demonstrated
by molecular dynamics (MD) simulations, how a mechanism involving overcharged
and undercharged spherical macroions could lead to a \textit{strong long range}
attraction between charged spheres. In this paper we give a more detailed account
and elaborate on the physical mechanism responsible for \textit{charge inversion}
(overcharge). Why and how does a charged particle strongly {}``bind{}'' electrostatically
at its surface so many counterions that its net charge changes sign? We further
will discuss the necessary ingredients to explain this phenomenon in terms of
a simple Wigner crystal theory. Using this Ansatz we show that it is possible
for a pair of colloids which are sufficiently different in charge density to
have an ionized ground state. Both, the one and two colloid cases, are treated
in terms of analytical predictions and verifications by simulation. Of special
interest are the energy barriers necessary to cross from a neutral pair to an
ionized pair state. We finally demonstrate by explicit simulations that the
described features survive also at finite temperature case.

The paper is organized as follows: in Sec. II a simple model based on the Gillespie
rule is proposed to understand charge inversion. Section III contains details
of our MD simulation model. Section IV is devoted for the study of a single
highly charged colloid. In Sec. V we investigate the situation where two colloids
are present. Finally, in Sec. VI we outline a summary of the results.

\section{Understanding overcharging via the Gillespie rule\label{sec.gillepsie}}

Here we propose a simple model solely based on electrostatic energy considerations
in order to understand the phenomenon of charge inversion for strongly coupled
systems. Because of the analogy between a spherical macroion surrounded by counterions
and an atom {[}i. e. nucleus + electrons{]}, it turns out be fruitful to use
classical pictures of atomic physics in order to gain comprehension of certain
phenomena occurring in mesoscopic colloidal systems \cite{Messina_PRL_2000,Messina_EPL_2000}.
To study the possibility of overcharging a single macroion, we recall the Gillespie
rule also known as the valence-shell electron-pair repulsion (VSEPR) theory
\cite{NOTE_VESPR_theory,Gillespie_JCE_1963} which is well
known in chemistry to predict the molecular geometry in covalent compounds.
Note that originally this model has nothing to do with overcharge. Applying
simple electrostatics one can compute that the \textit{ground state structure}
of two, three, four and five electrons disposed on a hard sphere corresponds
to simple geometrical situations like those depicted in Fig. \ref{fig.gillespie}.
The electrons try to maximize their mutual distances which leads, for example,
in the case of 3 and 4 electrons to equilateral triangular and tetrahedral arrangements.

Now, we can apply this concept to a spherical colloid of radius \( a \), central
charge \( Z_{m}=+2e \), where \textit{e} is the elementary charge, and \( N_{c} \)
monovalent counterions. By referring to Fig. \ref{fig.gillespie}, the \textit{neutral}
system corresponds to the case where two counterions are present, and the three
other cases (three, four and five counterions) correspond to \textit{non-neutral
overcharged} states.

The total electrostatic energy \( E(N_{c}) \) is merely made up of two terms:
i) an attractive term \( E_{att}(N_{c}) \) due to the attraction between the
counterions and the central charge and ii) a repulsive term \( E_{rep}(N_{c}) \)
due to the repulsion among the counterions. The final expression for the electrostatic
energy as a function of the number of counter ions reads 

\begin{equation}
\label{Eq.gillespie}
E(N_{c})=E_{att}(N_{c})+E_{rep}(N_{c})=k_{B}T\frac{l_{B}}{a}\left[ -N_{c}Z_{m}+f(\theta )\right] ,
\end{equation}
where \( l_{B}=e^{2}/(4\pi \epsilon _{0}\epsilon _{r}k_{B}T) \) is the
Bjerrum length and \( f(\theta ) \)
is the repulsive energy part which is solely a function of the \textit{topology}
(relative angles between counterions, such as \( \alpha  \) and \( \beta  \)
appearing in Fig. \ref{fig.gillespie}, which also depend on \textit{\( N_{c} \)}{]}
of the ground state figure. For the specific cases reported in Fig. \ref{fig.gillespie},
the calculation of \( E(N_{c}) \), with \( 2\leq N_{c}\leq 5, \) is straightforward
and the corresponding energy values are given in Fig. \ref{fig.gillespie}.
One deduces that the maximally obtainable overcharging is -2\textit{e} (i. e.
100\%) around the central charge. That is, the excess counterions gain more
energy by assuming a topological favorable configuration than by escaping to
infinity, the simple reason of overcharge. Note the arguments for overcharging
are independent of the Bjerrum length and of the sphere radius, which enter
only as prefactors in Eq. (\ref{Eq.gillespie}).

To safely use this above outlined model one has just to ensure that the counterion
size is small enough to avoid excluded volume effects, which in practice is
always true. The important message is that, from an energy point of view, a
colloid \textit{always} tends to be overcharged. Obviously, for high central
charge, the direct computation of the electrostatic energy by using the exact
equation (\ref{Eq.gillespie}) becomes extremely complicated. Therefore we resort
to simulations for highly charged spheres.

\section{Simulation model\label{sec.simulation-model}}

The system under consideration contains two types of spherical charges: (i)
one or two macroion(s) with a bare central charge \( Q=-Z_{m}e \) (with \( Z_{m}>0 \))
and (ii) small counterions of diameter \( \sigma  \) with charge \( q=+Z_{c}e \)
(with \textit{\( Z_{c}=2 \)}) to neutralize the whole system. All these ions
are confined in an impermeable cell and the macroion(s) is (are) held fixed.

The molecular dynamics (MD) technique employed here is similar to this used
in previous studies \cite{Messina_PRL_2000,Messina_EPL_2000}. In order to simulate
a canonical ensemble, the motion of the counterions is coupled to a heat bath
acting through a weak stochastic force \textbf{W}(t). The equation of motion
of counterion \textit{i} reads 

\begin{equation}
\label{eq. Langevin}
m\frac{d^{2}\mathbf{r}_{i}}{dt^{2}}=-\nabla _{i}U-m\gamma \frac{d\mathbf{r}_{i}}{dt}+{\mathbf{W}_{i}}(t),
\end{equation}
where \textit{m} is the counterion mass, \textit{U} is the potential force
having two contributions: the Coulomb interaction and the excluded volume interaction,
and \( \gamma  \) is the friction coefficient. Friction and stochastic force
are linked by the dissipation-fluctuation theorem \( <{\mathbf{W}_{i}}(t)\cdot {\mathbf{W}_{j}}(t')>=6m\gamma k_{B}T\delta _{ij}\delta (t-t^{'}) \).
For the ground state simulations the fluctuation force is set to zero.

Excluded volume interactions are taken into account with a purely repulsive
Lennard-Jones potential given by 

\begin{equation}
\label{eq. LJ}
U_{LJ}(r)=\left\{ \begin{array}{l}
4\epsilon _{LJ}\left[ \left( \frac{\sigma }{r-r_{0}}\right) ^{12}-\left( \frac{\sigma }{r-r_{0}}\right) ^{6}\right] +\epsilon _{LJ},\\
0,
\end{array}\right. \begin{array}{l}
\textrm{for}\, \, r-r_{0}<r_{cut},\\
\textrm{for}\, \, r-r_{0}\geq r_{cut},
\end{array}
\end{equation}
where \( r_{0}=0 \) for the counterion-counterion interaction, \( r_{0}=7\sigma  \)
for the macroion-counterion interaction, \( r_{cut} \) (= \( 2^{1/6}\sigma  \))
is the cutoff radius. This leads to an \textit{effective} macroion radius \( a \)
(\( a=r_{0}+\sigma =8\sigma  \)) corresponding physically to the macroion-counterion
distance of closest approach. Energy and length units in our simulations are
defined as \( \epsilon _{LJ}= \)\( k_{B}T_{0} \) (with \( T_{0}=298 \) K)
and \( \sigma =3.57 \) \AA\  respectively. In the following we will set \( k_{B}T_{0}=1 \),
so that all energies are measured in those units, suppressing thereby all factors
of \( k_{B}T_{0} \) in our equations.

The pair electrostatic interaction between any pair \textit{ij}, where \textit{i}
and \textit{j} denote either a macroion or a counterion, reads

 \begin{equation}
\label{eq.coulomb}
U_{coul}(r)=l_{B}\frac{Z_{i}Z_{j}}{r}\: ,
\end{equation}
where \( Z_{i} \) represents the valence of the ions (counterion or macroion).
Being essentially interested in the strong Coulomb coupling regime we choose
the relative permittivity \( \epsilon _{r}=16 \), corresponding to a Bjerrum
length of \( 10\sigma  \), for the remaining of this paper. To avoid image
charges complications, the permittivity \( \epsilon _{r} \) is supposed to
be identical within whole the cell (including the macroion) as well as outside
the cell. Typical simulation parameters are gathered in Table \ref{tab.simu-param}.

\section{One macroion case\label{sec.one-col}}

In this section, we focus on counterion distribution exclusively governed by
\textit{energy minimization}, i. e. \textit{T} = 0K. The single spherical macroion
is fixed to the center of the large outer spherical simulation cell (i. e. both
spheres are concentric) of radius $R=40\sigma$. This leads to a colloid volume fraction
$f_{m}={a^3}/{R^3}=8\times10^{-3}$.
In such a case correlations are maximal, and all the
counterions lie on the surface of the spherical macroion. To avoid being trapped
in metastable states, we systematically heated and cooled (10 cycles) the system
and only kept the lowest energy state then obtained \cite{NOTE_MD_barrier}.
It turns out that for this type of repulsive potential (between counterions)
no rough energy landscape appears and thus, the MD method is efficient to find
the ground state. First, we checked that this method reproduces well the ground
state energies and structures of the simple situations depicted in Fig. \ref{fig.gillespie}.

\subsection{Counterion distribution\label{sec.coun-distribution}}

To characterize the counterion layer \textit{structure}, we compute the counterion
correlation function \( g(r) \) on the surface of the sphere, defined as: \begin{equation}
\label{Eq. CCF-g(r)}
c^{2}g(r)=\sum _{i\neq j}\delta (r-r_{i})\delta (r-r_{j}),
\end{equation}
 where \( c=N/4\pi a^{2} \) is the surface counterion concentration (\( N \)
being the number of counterions), \textit{r} corresponds to the arc length on
the sphere. Note that at zero temperature all equilibrium configurations are
identical, thus only one is required to obtain \( g(r) \). The pair distribution
\( g(r) \) is normalized as follows 

\begin{equation}
\label{eq. correlation-normalization}
c\int _{0}^{\pi a}2\pi rg(r)dr=(N_{c}+n-1),
\end{equation}
where \( N_{c}=Z_{m}/Z_{c} \) is the number of counterions in the neutral
state and \textit{\( n \)} is the number of overcharging counterions. Because
of the \textit{finite} size and the topology of the sphere, \( g(r) \) has
a cut-off at \( \pi a \) (=25.1 \( \sigma  \)) and a \textit{zero} value there.
More precisely one cannot state that the uncorrelated case corresponds to \( g(r)=1 \)
for the present finite system. Therefore at {}``large{}'' distance the correlation
function differs from the one obtained with an infinite planar object. Furthermore
the absolute value of \( g(r) \) cannot be directly compared to the one obtained
with an infinite plane.

Correlation functions for the structural charge \( Z_{m}=180 \) and for two
states of charge, neutral (\( n=0 \)) and overcharged (\( n=8 \)), can be
inspected in Fig. \ref{fig.ground-state-CCF}. One remarks that both structures
are very similar and highly ordered. A snapshot of the ground state structure
of the neutral state (\( n=0 \)) is depicted in Fig. \ref{fig.gs-snapshot-OC_0}.
A visual inspection gives an almost perfect \textit{triangular} crystalline
structure (see Fig. \ref{fig.gs-snapshot-OC_0}). A closer look at Fig. \ref{fig.ground-state-CCF}
reveals that the \( g(r) \) of the overcharged state, containing eight more
counterions than the neutral one, shows its first peak at some shorter distance
compared to the \( g(r) \) of the neutral state, as is expected for denser
systems.

It is also interesting to know how the counterion-layer structure looks like
when the system is brought to \textit{room temperature} \( T_{0} \). At non
zero temperature, correlation functions are computed by averaging 
\( \sum _{i\neq j}\delta (r-r_{i})\delta (r-r_{j}) \)
over 1000 independent equilibrium configurations which are statistically uncorrelated.
Results are depicted in Fig. \ref{fig.temperature-CCF} for \( Z_{m}=180 \)
and \( f_{m}=8\times 10^{-3} \). As expected the long-range counterion positional
order is neatly weaker at room temperature than in the ground state case. Meanwhile,
the structure remains very correlated and highly short-range ordered and therefore
it is referred as a strongly correlated liquid (SCL) \cite{Shklowskii_PRE_1999b}.
In terms of Coulomb coupling parameter \cite{Shklowskii_PRE_1999b,Rouzina_JCP_1996}
$\Gamma =Z_{c}^{2}l_{B} /a_{cc} $ , where \( a_{cc} \) is the average distance between
counterions, we have \( \Gamma \approx 13 \) for \( Z_{m}=180 \).

\subsection{Energy analysis\label{sec.nrj-analysis}}

As demonstrated in Sec. II, the spatial correlations are fundamental to obtain
overcharge. Indeed, if we apply the same procedure and smear \( Z \) counterions
onto the surface of the colloid of radius \( a \), we obtain for the energy

\begin{equation}
\label{smeared}
E=l_{B}\left[ \frac{1}{2}\frac{Z^{2}}{a}-\frac{Z_{m}Z}{a}\right] .
\end{equation}
 The minimum is reached for \( Z=Z_{m} \), hence no overcharging occurs.

To generalize results of Sec. II to higher central charges we have considered
three macroionic charge \( Z_{m} \) of values 50, 90 and 180 corresponding
to a surface charge density of one elementary charge per 180, 100 and 50 \AA\( ^{2} \),
respectively. For a given macroion, we always start by adding the exact number
of counterions \( N_{c} \) to have an electro-neutral system. Once equilibrium
of this system is reached, we add the first overcharging counterion and let
the new non-neutral system relax, and we repeat this operation a given number
of times. The electrostatic energy is computed by summing up the pairwise interactions
of Eq. (\ref{eq.coulomb}) over all pairs.

The electrostatic energy as a function of the number of overcharging counterions
\( n \) is displayed Fig. \ref{fig.OC-MD-energy}. We note that the maximal
(critical) acceptance of \( n \) (4, 6 and 8) increases with the macroionic
charge \( Z_{m} \) (50, 90 and 180 respectively). Furthermore for fixed \( n \),
the gain in energy is always increasing with \( Z_{m} \). Also, for a given
macroionic charge, the gain in energy between two successive overcharged states
is decreasing with \( n \).

The results of Sec. \ref{sec.coun-distribution} showed that in the ground
state the counterions were highly ordered. Rouzina and Bloomfield
\cite{Rouzina_JCP_1996} first stressed the special importance of these
crystalline arrays for interactions of multivalent ions with DNA strands, and
later Shklovskii (\cite{Shklowskii_PRL_1999a,Shklowskii_PRE_1999b} and
references therein) showed that the Wigner crystal (WC) theory can be applied
to determine the interactions in strongly correlated systems. In two recent
short contributions \cite{Messina_PRL_2000,Messina_EPL_2000} we showed that
the overcharging curves obtained by simulations of the ground state, like Fig.
\ref{fig.OC-MD-energy}, can be simply explained by assuming that the energy \(
\varepsilon \) per counterion on the surface of a macroion depends linearly on
the inverse distance between them, hence is proportional to \( \sqrt{N} \) for
fixed macroion area, where \( N \) is the \textit{total} number of counterions
on the surface \cite{Messina_PRL_2000,Messina_EPL_2000,note_PRE_WC_curvature}.
This can be justified by the WC theory. The idea is that the counterions form
an ordered lattice on the surface of a homogeneously charged background of
opposite charge, which is also called a One Component Plasma
(OCP)\cite{ocp_ref}. Each ion interacts in first approximation only with the
oppositely charged background of its Wigner-Seitz (WS)
cell\cite{Shklowskii_PRE_1999b}, which can be approximated by a disk of radius
\( h \), which possesses the same area as the WS cell. Because we can assume
the area of the WS cell to be evenly distributed among the \( N \) counterions
on the sphere's surface \( A=4\pi a^{2} \) we find

\begin{equation}
\label{eq:h}
\pi h^{2}=\frac{A}{N}=c^{-1}.
\end{equation}
The electrostatic interaction energy \( \varepsilon ^{(h)} \) of one counterion
with the background of its WS cell can then be determined by 

\begin{equation}
\label{Eq.hole}
\epsilon ^{(h)}=-l_{B}Z_{c}^{2}\int _{0}^{h}2\pi rc\frac{1}{r}dr=-2\sqrt{\pi }l_{B}Z^{2}_{c}\sqrt{c},
\end{equation}
hence is proportional to \( \sqrt{c} \), which proves our initial assumption.
It is convenient to define \( \ell =l_{B}Z_{c}^{2} \) and \( \alpha ^{(h)}=2\sqrt{\pi }\approx 3.54 \).
For fixed macroion area we can then rewrite Eq. (\ref{Eq.hole}) as 

\begin{equation}
\label{eq.E-N}
\varepsilon ^{(h)}(N)=-\frac{\alpha ^{(h)}\ell }{\sqrt{A}}\sqrt{N}.
\end{equation}
If one computes this value for an infinite plane, where the counterions form
an exact triangular lattice, and takes into account all interactions, one
obtains the same form as in Eq. (\ref{Eq.hole}), but the prefactor \( \alpha
^{(h)} \) gets replaced by the numerical value \( \alpha ^{WC}=1.96
\)\cite{Bonsall_PRB_1977}.  Although the \textit{value} is almost a factor of
two smaller than the simple hole picture suggests, the \textit{functional
  dependence} on the concentration is still the same.

Not knowing the precise value of \( \alpha \) we can still use the simple
scaling behavior with \( c \) to set up an equation to quantify the energy
gain \( \Delta E_{1} \) by adding the first overcharging counterion to the
colloid. To keep the OCP neutral we imagine adding a homogeneous surface
charge density of opposite charge ($\frac{-Z_c e}{A}$) to the
colloid\cite{houches}. This 
ensures that the background still neutralizes the incoming overcharging
counterion and we can apply Eq. (\ref{eq.E-N}). To cancel our surface charge
addition we add another homogeneous surface charge density of opposite sign
$\frac{Z_c e}{A}$. This surface charge does not interact with the now neutral
OCP, but adds a self-energy term of magnitude $\frac{1}{2}\frac{\ell}{a}$, so
that the total energy difference for the first overcharging counterion reads as
 \begin{equation}
\label{Eq.WC-FIRST-OC}
\Delta E_{1}=(N_{c}+1)\varepsilon (N_{c}+1)-N_{c}\varepsilon (N_{c}) + \frac{\ell}{2a}.
\end{equation}
By using Eq. (\ref{eq.E-N}) this can be rewritten as 

\begin{equation}
\label{Eq.WC-FIST-OC-b}
\Delta E_{1}=-\frac{\alpha \ell }{\sqrt{A}}\left[
  (N_{c}+1)^{3/2}-N_{c}^{3/2}\right] + \frac{\ell}{2a} .
\end{equation}
Completely analogously one derives for
the energy gain \( \Delta E_{n} \) for \( n \) overcharging counterions\cite{note}
\begin{equation}
\label{Eq.WC-n-OC}
\Delta E_{n}=-\frac{\alpha \ell }{\sqrt{A}}\left[ (N_{c}+n)^{3/2}-N_{c}^{3/2}\right] +\frac{\ell }{a}\frac{n^2}{2}.
\end{equation}
Equation (\ref{Eq.WC-n-OC}) can be seen as an approximation of the exact general
expression Eq. (\ref{Eq.gillespie}), where the topological term \( f(\theta ) \)
is handled by assuming a perfect planar crystalline structure through Eqs. 
(\ref{Eq.WC-FIRST-OC}-\ref{Eq.WC-n-OC}).
Using Eq. (\ref{Eq.WC-n-OC}), where we determined the unknown \( \alpha  \)
from the simulation data for \( \Delta E_{1} \) via Eq. (\ref{Eq.WC-FIST-OC-b})
we obtain a curve that matches the simulation data almost perfectly, compare
Fig. \ref{fig.OC-MD-energy}. The second term in Equation (\ref{Eq.WC-n-OC})
also shows why the overcharging curves of Fig. \ref{fig.OC-MD-energy} are shaped
parabolically upwards for larger values of \( n \).

Using the measured value of \( \alpha  \) we can simply determine the maximally
obtainable number \( n_{max} \) of overcharging counterions by finding the
stationary point of Eq. (\ref{Eq.WC-n-OC}) with respect to \( n \):

\begin{equation}
\label{Eq.Q*}
n_{max}=\frac{9\alpha ^{2}}{32\pi }+
\frac{3\alpha }{4\sqrt{\pi }}\sqrt{N_{c}}
\left[ 1+ \frac{9\alpha^2}{64\pi N_c }\right] ^{1/2}.
\end{equation}
The value of \( n_{max} \) depends only on the number of counterions $N_{c}$
 and \( \alpha \). For large \( N_{c} \) Eq. (\ref{Eq.Q*}) reduces to \(
n_{max}\approx \frac{3\alpha }{4\sqrt{\pi }}\sqrt{N_{c}} \) which was derived
in Ref. \cite{Shklowskii_PRE_1999b} as the low temperature limit of a a
neutral system in the presence of salt. What we have shown is that the
overcharging in this limit has a pure electrostatic origin, namely it
originates from the topological favorable arrangement of the ions around a
central charge.  In the following we will investigate the behavior of \(
\alpha \) on the surface charge density and on the radius of the macroion.

We have performed simulations for various surface charge densities by keeping
\( A \) fixed and changing \( Z_{m}=2N_{c} \) in the range 2 up to 180.
Results can be found in Table \ref{tab.WC} and in Fig. \ref{fig.alpha-WC-N}.
We observe that \( \alpha \) is already for values of \( N_{c} \) as small as
two, where one can use the Gillespie rule to calculate the energy exactly,
close the planar value $\alpha^{(WC)}$, and actually oscillates around this
value.  For $N_C > 50$, one reaches a plateau of \( \alpha =1.86\pm 0.05 \).

This value is about 5\% smaller then the one predicted by WC theory, and is
presumably due to the finite curvature of the sphere. For large values of the
radius \( a \) we expect \( \alpha \) to reach the planar limit. To see the
rate of convergence we varied\footnote{%
  Note that this is the only part of the paper where \( a\neq 8\sigma \).  }
$a$ at a fixed concentration $c$. The results can be found in Table
\ref{tab.WC-curvature} and Fig. \ref{fig.alpha-WC-a}. For our smallest value
of \( a=6\sigma \) we find \( \alpha =1.91 \). For small $a$, which is
equivalent to a small number of $N_c$, we observe again a slight oscillatory
behavior of $\alpha$, whereas for our two largest
values $a=80\sigma$ and $160\sigma$ we find up to numerical uncertainties the planar
result \( \alpha = \alpha ^{WC}=1.96 \).  Again we stress that the
numerical value of \( \alpha \) enters only as a prefactor into the equations
which govern the overcharging, it does not change the qualitative behavior.

One could wonder if the results presented above are still valid when the bare
central charge of the colloid is replaced by small \textit{discrete} ions lying
on the macroion surface? In fact it has been shown that the energy of the overcharged
state (Fig. \ref{fig.OC-MD-energy}) for \textit{random} discrete colloidal
charge distribution is more or less quantitatively affected \cite{Messina_EPJE_2001,Messina_dcc2_2001}
depending on the the valence of the counterions. More precisely it was shown
that the overcharge still persists and has a similar (for monovalent counterions)
or quasi-identical (for multivalent counterions) behavior to the one depicted
in Fig. \ref{fig.OC-MD-energy}, and this, even if \textit{ionic pairing} occurs
between the counterions and the discrete colloidal charges \cite{Messina_EPJE_2001,Messina_dcc2_2001},
that is even when \textit{no} counterion WC is formed.

\subsection{Macroion-counterion interaction profile\label{sec.Interaction-profile}}

In this part, we study the interaction potential profile at \( T=0K \) between
a \textit{neutral effective} macroion {[}bare macroion + neutralizing counterions{]}
and one excess overcharging counterion at a distance \( r \) from the colloid
center. The profile is obtained by displacing adiabatically the excess overcharging
counterion from infinity towards the macroion. We investigated the case of 
$Z_{m}=2,4,6,8,10,32,50,90,128,180$,
and \( 288 \). All curves can be nicely fitted with an exponential fit of the
form 

\begin{equation}
\label{eq.Interaction-fit}
E_{1}(r)=\Delta E_{1}e^{-\tau (r-a)},
\end{equation}
where \( \Delta E_{1} \) is the measured value for the first overcharging
counterion, and \( \tau  \) is the only fit parameter (see Table \ref{tab.WC}).
Results for the two values \( Z_{m}=50 \) and \( 180 \) are depicted in Fig.
\ref{fig.gs-interaction-PROFILE}. If one plots all our results for \( \tau  \)
versus \( \sqrt{N_{c}} \) we observe a linear dependence for a wide range of
values for \( N_{c} \), 

\begin{equation}
\label{eq.tau}
\tau =m\sqrt{N_{c}},
\end{equation}
 with \( m\sigma \approx 0.1 \), as can be inspected in Fig. \ref{fig.tau-WC}.

This behavior can again be explained using a {}``WC hole{}'' picture in the
limiting situation where \( x:=r-a \) is small (i. e. the displaced counterion
is close to the macroion surface). To this end we consider the classical electrostatic
interaction \( V_{disk}(x) \) between a uniformly charged disk (the WC hole
- supposed planar) and a point ion (the displaced counterion) located on the
axis of the disk at a distance \( x \) from its surface, which is given by

\begin{equation}
\label{eq.disk-interaction}
V_{disk}(x)=-2\pi \ell c(\sqrt{h^{2}+x^{2}}-x).
\end{equation}
 As in Eq. (\ref{Eq.hole}), \( h=(\pi c)^{-1/2} \) is the hole radius. For
small distance \( x \), we expand Eq. (\ref{eq.disk-interaction}) \begin{equation}
\label{eq.disk-approx}
V_{disk}(x)=\varepsilon ^{(h)}\left[ 1-\frac{1}{h}x+
\frac{1}{2h^{2}}x^{2}+{\mathcal{O}}\left( \frac{x^{4}}{h^{4}}\right) \right] ,
\end{equation}
 where the surface term \( V_{disk}(x=0)=\varepsilon ^{(h)} \) is given by
Eq. (\ref{Eq.hole}). By expanding the exponential in Eq. (\ref{eq.Interaction-fit})
to \( 2^{nd} \) order for small \( \tau x \) we obtain 

\begin{equation}
\label{eq.Interaction-fit-linear}
E_{1}(x)=\Delta E_{1}\left[ 1-\tau x+\frac{\tau ^{2}}{2}x^{2}+{\mathcal{O}}(\tau ^{3}x^{3})\right] .
\end{equation}

A comparison between Eq. (\ref{eq.Interaction-fit-linear}) and Eq. (\ref{eq.disk-approx})
shows that to this order we can identify 

\begin{equation}
\label{eq.tau-h}
\tau =\frac{1}{h}=\sqrt{\pi c}=\frac{\sqrt{N_{c}}}{2a}\approx 0.06\sqrt{N_{c}}.
\end{equation}
Comparing this to Eq. (\ref{eq.tau}) we note that this simple illustration
gives us already the correct scaling as well as the prefactor up to 30\%. We
neglected here the effect that the surface concentration changes when the ion
is close to the macroion as well as the curvature of the macroion.

\section{Two macroions case\label{sec.two-col}}

In this section we consider two fixed charged spheres of bare charge \( Q_{A} \)
and \( Q_{B} \) separated by a center-center separation \textit{\( R \)} and
surrounded by their neutralizing counterions\textit{.} All these ions making
up the system are immersed in a cubic box of length \( L=80\sigma  \), and
the two macroions are held fixed and disposed symmetrically along the axis passing
by the two centers of opposite faces. This leads to a colloid volume fraction
\( f_{m}=2\cdot \frac{4}{3}\pi (a/L)^{3}\approx 8.4\times 10^{-3} \). For \textit{finite}
colloidal volume fraction \( f_{m} \) and temperature, we know from the study
carried out above that in the strong Coulomb coupling regime all counterions
are located in a spherical {}``monolayer{}'' in contact with the macroion.
Here, we investigate the mechanism of \textit{strong long range} attraction
stemming from \textit{monopole} contributions: that is one colloid is overcharged
and the other one undercharged.

\subsection{Like charged colloids\label{sec.like-charged}}

\subsubsection{Observation of metastable ionized states\label{sec.obsevation-metastable-states}}

In the present charge symmetrical situation we have \( Q_{A}=Q_{B}=-Z_{m}e \).
This system is brought at \textit{room temperature} \textit{\( T_{0} \)}. Initially
the counterions are randomly generated inside the box. Figure \ref{fig.snapshot-179-181}
shows two macroions of bare charge \( Z_{m}=180 \) surrounded by their quasi-two-dimensional
counterions layer. The striking peculiarity in this configuration is that it
corresponds to an overcharged and an undercharged sphere. There is one counterion
more on the left sphere and one less on the right sphere compared to the bare
colloid charge. Such a configuration is referred as \textit{ionized state}.
In a total of 10 typical runs, we observe this phenomenon 5 times. We have also
carefully checked against a situation with periodic boundary conditions, yielding
identical results. However it is clear that such a state is {}``metastable{}''
because it is not the lowest energy state. Indeed, in this symmetrical situation
the ground state should also be symmetrical so that both colloid should be exactly
charge-compensated. Such arguments remain valid even at non-zero temperature
as long as the system is strongly energy dominated, which is presently the case.
Nevertheless the ionized states observed here seem to have a long life time
since even after \( 10^{8} \) MD time steps this state survives. In fact we
could not observe within the actual computation power the recover of the stable
neutral state. To understand this phenomenon we are going to estimate the energy
barrier involved in such a process.

\subsubsection{Energy barrier and metastability\label{sec..nrrj-metastability}}

To estimate the energy barrier, electrostatic energy profiles at \textit{zero
temperature} were computed, where we move one counterion from the overcharged
macroion to the undercharged, restoring the neutral state {[}see drawing depicted
in Fig. \ref{fig.barrier}(a){]}. We have checked that the path leading to the
lowest barrier of such a process corresponds to the line joining the two macroions
centers. The simulation data are sketched in Figs. \ref{fig.barrier}(a-b) and
were fitted using a similar technique to the single macroion-counterion interaction
profile given by Eq. (\ref{eq.Interaction-fit}), which will be explicitly treated
later. The resulting simulated energy barrier \( \Delta E_{bar} \) is obtained
by taking the difference between the highest energy value of the profile and
the ionized state energy (start configuration). Values of \( \Delta E_{bar} \)
can be found in Table \ref{tab.fit-small-R} for the small macroion separation
case \( R/a=2.4 \). One clearly observes a barrier, which increases quasi linearly
with the charge \textit{\( Z_{m} \)} for the small colloids separation \( R/a=2.4 \)
{[}cf. Fig. \ref{fig.barrier}(a) and Table \ref{tab.fit-small-R}{]}. The ground
state corresponds as expected to the neutral state. Note that the ionized state
and the neutral state are separated by only a small energy amount (less than
\( 2.5 \)), the difference being approximately of the order of the monopole
contribution \( E=l_{B}(4/8-4/11)\approx 1.36 \). The physical origin of this
barrier can be understood from the single macroion case where we showed that
a counterion gains high correlational energy near the surface. This gain is
roughly equal for both macroion surfaces and decreases rapidly with increasing
distance from the surfaces, leading to the energy barrier with its maximum near
the midpoint. For the single macroion case we showed that the correlational
energy gain scales with \( \sqrt{Z_{m}} \), whereas here we observe a linear
behavior of the barrier height with \( Z_{m} \). We attribute this effect to
additional ionic correlations since both macroions are close enough for their
surface ions to interact strongly. For large separations (here \( R/a=4.25 \))
we find again that the barrier height increases with \( \sqrt{Z_{m}} \), as
expected {[}see Fig. \ref{fig.barrier}(b) and Table \ref{tab.fit-large-R}{]}.
Furthermore the energy barrier height naturally increases with larger colloidal
separation. The \( Z_{m} \) dependence of the barrier also shows that at room
temperature such ionized states only can occur for large \( Z_{m} \). In our
case only for \textit{\( Z_{m}=180 \)}, the ionized state was stable for all
accessible computation times. Unfortunately, it is not possible to get a satisfactory
accuracy of the energy jumps at non-zero temperatures. Nevertheless, since we
are interested in the strong Coulomb coupling regime, which is energy dominated,
the zero temperature analysis is sufficient to capture the essential physics.

Simulation results presented in Fig. \ref{fig.barrier} can be again theoretically
well described using the previously exponential profiles obtained for the macroion-displaced
counterion in Sec. \ref{sec.Interaction-profile} for a \textit{single} colloid.
For the two macroions case, the general expression for the electrostatic interaction
\( E_{bar}(r,R) \) of the present process can be approximated as 

\begin{equation}
\label{eq.barrier-fit-2col}
E_{bar}(r,R)=\Delta E^{*}_{1}\exp \left[ -\tau (r-a)\right] 
+\Delta E^{*}_{1}\exp \left[ -\tau (R-r-a)\right] - \frac{\ell }{R-r},
\end{equation}
where \( \Delta E^{*}_{1} \) is the {}``effective{}'' \textit{correlational}
energy gained by the first OC at one macroion surface assumed identical for
both colloids. The last term in Eq. (\ref{eq.barrier-fit-2col}) corresponds
to the additional monopole attractive contribution of the displaced counterion
with the undercharged colloid. Fitting parameters (\( \Delta E^{*}_{1} \) and
\( \tau  \)) for \( R/a=2.4 \) and \( R/a=4.25 \) can be found in Tables
\ref{tab.fit-small-R} and \ref{tab.fit-large-R} respectively. Same values
of \( \tau  \) were used here as those of the single macroion case (see Fig.
\ref{fig.tau-WC} and Table \ref{tab.WC}) . However for the small colloidal
separation (\( R/a=2.4 \)), due to the extra inter-colloidal surface counterions
correlations, we used a slightly larger (absolute) value for \( \Delta E^{*}_{1} \)
compared to the one (\( \Delta E_{1} \)) of an isolated colloid (compare Tables
\ref{tab.fit-small-R} and \ref{tab.fit-large-R}). This is compatible with
the idea that between the two colloids (especially when both spheres come at
contact), we have the formation of a {}``super-layer{}'' which is more dense,
thus leading to a smaller hole radius and a higher energy gain. An analysis
of the counterions structure of the two macroions reveals that both WC counterion
layers are interlocked, that is the projection along the axis passing through
the colloid centers gives a superlattice structure (see Fig. \ref{fig.superlattice}).

For large colloidal separation (\( R/a=4.25 \)), the WC structure on one of
the colloids is unperturbed by the presence of the other, hence we can take
\( \Delta E^{*}_{1}=\Delta E_{1} \), and our simulation data can nicely be
fitted by the parameters inferred from the single colloid system.

\subsubsection{Effective forces}

Results concerning the effective forces at \textit{zero temperature} between
the two macroions are now investigated which expression is given by 

\begin{equation}
\label{Eq.effective-force}
F_{eff}(R)=F_{mm}(R)+F_{LJ}+F_{mc},
\end{equation}
 where \( F_{mm}(R) \) is the direct Coulomb force between macroions, \( F_{LJ} \)
is the excluded volume force between a given macroion and its surrounding counterions
and \( F_{mc} \) is the Coulomb force between a given macroion and all the
counterions. Because of symmetry, we focus on one macroion. To understand the
extra-attraction effect of these ionized-like states, we consider three cases:
(i) \( F_{ion}=F_{eff} \) in the ionized state with a charge asymmetry of \( \pm  \)
1 counterion (ii) \( F_{neut}=F_{eff} \) in the neutral case (iii) \( F_{mono}=F_{eff} \)
simply from the effective monopole contribution. Our results are displayed in
Fig. \ref{fig.Effective-Force} for \( Z_{m}=180 \) , where the ionized state
was also observed at room temperature. The non-compensated case leads to a very
important extra attraction. This becomes drastic for the charge asymmetry of
\( \pm  \) 2 counterions at short separation \( R/a=2.4 \) leading to a reduced
effective attractive force \( Fl_{B}=-10.7 \), a situation which was also observed
in our simulation at room temperature. In contrast to previous studies 
\cite{Jensen_Physica_1998,Elshad_PRL_1998},
these attractions are long range. For a sufficiently large macroion separation
(from \( 3.5a \)), corresponding here roughly to a macroion surface-surface
separation of one colloid diameter, the effective force approaches in good approximation
the monopole contribution (see Fig. \ref{fig.Effective-Force}).

\subsection{Asymmetrically charged colloids}

In this section we investigate the case where the two colloids have different
charge densities. We will keep the colloidal radii \( a \) fixed, but vary the
bare colloidal charges. The charge on sphere \textit{A} is \textit{fixed} at
\( Z_{A}=180 \), and sphere \textit{B} carries \textit{variable} charges with
\( Z_{B} \) (where \( Z_{B}<Z_{A} \)) ranging from 30 up to 150. Global
electroneutrality is ensured by adding \( N_{A}+N_{B} \) divalent counterions,
with \( N_{A}=Z_{A}/Z_{c} \), and \( N_{B}=Z_{B}/Z_{c} \). In this way we vary
the bare counterion concentrations \( c_{i}=\sqrt{N_{i}/4\pi a^{2}} \), where
\( i \) stands for \( A \) or \( B \).

\subsubsection{Ground state analysis\label{sec.gs-2colAB}}

We start out again with studying the ground state of such a system. The
electrostatic energy of the system is investigated for different uncompensated
bare charge cases (ionized states) by simply summing up Eq. (\ref{eq.coulomb})
over all Coulomb pairs. We define the \textit{degree of ionization}
(\textit{DI}) as the number of counterions overcharging colloid A (or,
equivalently, undercharging colloid B). The system is prepared in various
\textit{DI} and we measure the respective energies. These states are separated
by kinetic energy barriers, as was demonstrated above. We consider three
typical macroionic charges \( Z_{B} \) (30, 90 and 150) and separations \( R/a
\) (2.4, 3.0 and 4.25). The main results of the present section are given in
Fig. \ref{fig.gs-DI}. For the largest separation \( R/a=4.25 \) and largest
charge \( Z_{B}=150 \) {[}see Fig. \ref{fig.gs-DI}(a){]}, one notices that the
ground state corresponds to the classical compensated bare charge situation
{[}referred as the \textit{neutral} \textit{state} (\textit{DI}=0){]}.
Moreover the energy increases stronger than linear with the degree of
ionization.  If one diminishes the bare charge \( Z_{B} \) to 90 and 30, the
\textit{ground state} is actually the ionized state for a \textit{DI} of 1 and
3, respectively.  The ionized ground state is about 8 and 36 , respectively,
lower in energy compared to the neutral state. This shows that even for a
relative large colloid separation, stable ionized states should exist for
sufficient low temperatures and that their stability is a function of their
charge asymmetry.

For a shorter separation \( R/a=3.0 \), ionized ground states are found {[}see
Fig. \ref{fig.gs-DI}(b){]} for the same charges \( Z_{B} \) as previously.
Nevertheless, in the ground state the \textit{DI} is now increased and it
corresponds to 2 and 4 for \( Z_{B}=90 \) and 30 respectively. The gain in
energy is also significantly enhanced. For the shortest separation under
consideration \( R/a=2.4 \) {[}see Fig. \ref{fig.gs-DI}(c){]}, the ground
state corresponds for \textit{all} investigated values of \( Z_{B} \) to the
ionized state, even for \( Z_{B}=150 \).  We conclude that decreasing the
macroion separation \textit{R} enhances the degree of ionization and the stability of
the ionized state.

To understand this ionization phenomenon, it is sufficient to refer to an
\textit{isolated} macroion surrounded by its neutralizing
counterions\textit{.} We have investigated the energies involved in the
ionization (taking out counterions). The complementary process of overcharging
(adding counterions) has already been investigated (see Fig.
\ref{fig.OC-MD-energy}). A derivation of the formula describing the ionization
energy \( \Delta E^{ion} \) proceeds completely analogously to the one carried
out for the overcharging Eq. (\ref{Eq.WC-n-OC}) and gives for the \( n^{th} \)
degree of ionization

\begin{equation}
\label{Eq.WC-Ionization}
\Delta E^{ion}_{n}=-
\frac{\alpha ^{B}\ell }{\sqrt{A}}
\left[ (N_B - n)^{3/2} - N_B^{3/2} \right] 
+ \frac{\ell }{a}\frac{n^2}{2},
\end{equation}
where $\alpha^{A,B}$ are the values of $\alpha$ belonging to colloid $A$ and $B$,
respectively. In Fig. \ref{fig.Ionization} we compare the predictions of Eqs.
(\ref{Eq.WC-n-OC}, \ref{Eq.WC-Ionization}) to our simulation data, which shows
excellent agreement.  Our numerical data for \( \Delta E^{ion}_{1} \) for \(
N_{B}=15 \), 45, and 75, the value of \( \Delta E^{OC}_{1} \) for \( N_{A}=90
\) (overcharging process), as well as the corresponding values for \( \alpha
\), which have been used for Fig. \ref{fig.Ionization} can be found in Table
\ref{tab.WC}.

With the help of Eqs. (\ref{Eq.WC-n-OC}, \ref{Eq.WC-Ionization}), one can
try to predict the curves of Fig. \ref{fig.gs-DI} for finite center-center
separation \textit{R}. Using for colloid \( A \) and \( B \) the measured
values \( \alpha ^{A} \) and \( \alpha ^{B} \), we obtain for the electrostatic
energy difference at finite center-center separation \( R \)

\begin{eqnarray}
\nonumber
\Delta E_{n}(R) & = & \Delta E^{ion}_{n}+\Delta E^{OC}_{n} \\
\nonumber & = &
\frac{3n\alpha ^{B}\ell }{2\sqrt{A}}\sqrt{N_{B}}
\left[1-\frac{n}{4N_{B}} 
+ {\mathcal{O}}\left( \frac{n^2}{N^{2}_{B}}\right)\right]
-\frac{3n\alpha ^{A}\ell }{2\sqrt{A}}\sqrt{N_{A}}
\left[1+\frac{n}{4N_{A}}
+ {\mathcal{O}}\left( \frac{n^2}{N^{2}_{A}}\right)\right] \\
& & + \frac{n^{2} \ell }{a}(1-\frac{a}{R}).\label{eq.finiteR}
\end{eqnarray}

The quality of the theoretical curves can be inspected in Fig.
\ref{fig.gs-DI}.  The prediction is is very good for large separations, but
the discrepancies become larger for smaller separations, and one observes that
the actual simulated energies are lower. Improvements could be achieved by
including polarization effects along the ideas leading to Eq. 
((\ref{eq.barrier-fit-2col}), by adjusting, for example, $\alpha^A$ and
$\alpha^B$. 
More important, the physical interpretation of Eq. (\ref{eq.finiteR}) is straightforward. The left
two terms represent the difference in correlation energy, and last term on the
right the monopole penalty due to the ionization and overcharging process.
This means that the correlational energy gained by overcharging the highly
charged colloid \textit{A} must overcome the loss of correlation energy as
well as the monopole contribution (\textit{two} penalties) involved in the
ionization of colloid \textit{B}. With the help of Eq. (\ref{eq.finiteR}) we
can establish a simple criterion (more specifically a sufficient condition),
valid for large macroionic separations, for the charge asymmetry \(
\sqrt{N_{A}}-\sqrt{N_{B}} \) to produce an ionized ground state of two unlike
charged colloids with the same size:

\begin{equation}
\label{Eq.criterion}
\left( \sqrt{N_{A}}-\sqrt{N_{B}}\right) >\frac{4\sqrt{\pi }}{3\alpha}.
\end{equation}
Referring to Fig. (\ref{fig.Ionization}) this criterion is met when the overcharge
curve (changed sign) is higher than the ionization curve.

If one uses the parameters of the present study one finds the requirement \( N_{B}<66 \)
to get a stable ionized state. This is consistent with our findings where we
show in Fig. \ref{fig.gs-DI} that for \( N_{B}=75 \)\textit{,} and \textit{R/a}
= 4.25, no ionized ground state exists whereas for \( N_{B}=60 \) we observed
one even for infinite separation (not reported here). The criterion Eq. (\ref{Eq.criterion})
is merely a sufficient condition, since we showed in Fig. \ref{fig.gs-DI} that
when the colloids are close enough this ionized state can appear even for smaller
macroion charge asymmetry due to enhanced inter-colloidal correlations. At this
stage, we would like to stress again, that the appearance of a stable ionized
ground state is due merely to correlation. An analogous consideration with smeared
out counterion distributions along the lines of Eq. (\ref{smeared}) will again
always lead to two colloids exactly neutralized by their counterions \cite{Schiessel_private_com}.
Our energetical arguments are quite different from the situation encountered
at finite temperatures, because in this case even a Poisson-Boltzmann description
would lead to an asymmetric counterion distribution. However, in the latter
case this happens due to pure entropic reasons, namely in the limit of high
temperatures, the counterions want to be evenly distributed in space, leading
to an effective charge asymmetry. 

At this stage, on looking at the results presented above, it appears natural
and straightforward to establish an analogy with the concept of ionic bonding.
It is well known in chemistry that the electro-negativity concept provides a
simple yet powerful way to predict the nature of the chemical bonding 
\cite{Pauling-Electronegtaivity-(1939)}.
If one refers to the original definition of the electro-negativity given by
Pauling \cite{Pauling-Electronegtaivity-(1939)}: {}``the power of an atom
in a molecule to attract electrons to itself{}'', the role of the bare charge
asymmetry becomes obvious. Indeed, it has an equivalent role at the mesoscopic
scale as the electron affinity at the microscopic scale. Another interesting
analogy is the influence of the colloidal separation on the stability of the
ionized state. Like in diatomic molecules, the ionized state will be (very)
stable only for sufficiently short colloid separations. Nevertheless, one should
not push this analogy too far. One point where it breaks down concerns the existence
of an ionized ground state in colloidal system for \textit{large} colloid separation,
providing that the difference in the counterion concentration on the surface
is large enough. In an atomistic system this is impossible since even for the
most favorable thermodynamical case, namely CsCl, there is a cost in energy
to transfer an electron from a cesium atom to a chlorine atom. Indeed, the smallest
existing ionization energy (for Cs, 376 kJ mol\( ^{-1} \)) is greater in magnitude
than the largest existing electron affinity (for Cs, 349 kJ mol\( ^{-1} \)).
In other terms, for atoms separated by large distances in the gas phase, electron
transfer to form ions is always energetically unfavorable.

\subsubsection{Finite temperature analysis}

As a last result, aimed at experimental verification, we show that an ionized
state can also exist \textit{spontaneously} at \textit{room temperature} \( T_{0} \).
Figure \ref{fig.relaxation} shows the time evolution of the electrostatic energy
of a system \( Z_{A}=180 \) with \( Z_{B}=30 \), \( R/a=2.4 \) and \( f_{m}=7\cdot 10^{-3} \),
where the starting configuration is the neutral state (\textit{DI} = 0). One
clearly observes two jumps in energy, \( \Delta E_{1}=-19.5\,  \) and \( \Delta E_{2}=-17.4\,  \),
which corresponds each to a counterion transfer from colloid \textit{B} to colloid
\textit{A}. These values are consistent with the ones obtained for the ground
state, which are\( -20.1\,  \) and \( -16.3\,  \) respectively. Note that
this ionized state (\textit{DI} = 2) is more stable than the neutral but is
expected to be metastable, since it was shown previously that the most stable
ground state corresponds to \textit{DI} = 5. The other stable ionized states
for higher \textit{DI} are not accessible with reasonable computer time because
of the high energy barrier made up of the correlational term and the monopole
term which increases with \textit{DI} . In Fig. \ref{fig.snaphot-DI2} we display
a typical snapshot of the ionized state (\textit{DI} = 2) of this system at
room temperature.

Obviously, these results are not expected by a DLVO theory even in the asymmetric
case (see e. g. \cite{DLVO_Assymety_DAguanno}). Previous simulations of asymmetric
(charge and size) spherical macroions \cite{Elshad_PRE_1998_Assymetric_Macroions}
were also far away to predict such a phenomenon since the Coulomb coupling was
weak (water, monovalent counterions).

\section{Concluding remarks\label{sec.conclusion}}

In summary, we have shown that the ground state of a charged sphere in the presence
of excess counterions is \textit{always} overcharged. A sufficiently charged
colloid can in principle be highly overcharged due to counterion mediated correlation
effects, and this phenomenon is quantitatively well described by a simple version
of Wigner crystal theory. In the strong Coulomb coupling regime, the energy
gain of a single excess ion close to a counterion layer can be of the order
of many tens of \( k_{B}T_{0} \). Furthermore we demonstrated that the electrostatic
interaction between a counterion and a macroion effectively neutralized by its
counterions decays exponentially on a length scale which is equal to the Wigner
crystal hole radius.

We further found that for two \textit{like-charged} macroions (symmetric case),
an initially randomly placed counterion cloud of their neutralizing divalent
counterions may not be equally distributed after relaxation, leading to two
macroions of opposite net charges. This is due to the short range WC attraction
which leads to this energetically favorable overcharged state. The resulting
configuration is metastable, however separated by an energy barrier of several
\( k_{B}T_{0} \) when the bare charge is sufficiently large, and can thus survive
for long times. Such configuration possess a natural strong long range attraction.

In return, if the symmetry in the counterion concentration on the colloidal
surface is sufficiently broken, the ionized state can be \textit{stable}. The
ground state of such a system is mainly governed by two important parameters,
namely the asymmetry in the counterion concentration determined by \( \sqrt{c_{A}}-\sqrt{c_{B}} \),
and the colloid separation \textit{R}. If the counterion concentration difference
is high enough, the ground state corresponds to an ionized state, whatever the
macroions separation \textit{\( R \)} is. However, the degree of ionization
depends on \textit{R}. Besides, for large \textit{R}, we have established a
criterion, allowing to predict when a stable ionized configuration can be expected.
The counterion concentration difference plays an analogous role to the electron
affinity between two atoms forming a molecule with ionic bonding. We demonstrated
that the results presented here for the ground state can lead to a stable ionic
state even at room temperature providing that the Coulomb coupling and/or the
counterion concentration asymmetry is sufficiently large. This is also a possible
mechanism which could lead to strong long range attractions, even in bulk. Future
work will treat the case where salt ions are present.

\acknowledgments

This work is supported by \textit{Laboratoires Europ\'{e}ens Associ\'{e}s}
(LEA) and a computer time grant hkf06 from NIC J\"{u}lich. We acknowledge helpful
discussions with B. J\"{o}nsson, R. Kjellander, H. Schiessel, and B. Shklovskii.


\begin{thebibliography}{10}
\bibitem{NOTE_MD_barrier}
Since we are using (MD) simulations, to avoid barrier energy problem the
  counterions were generated in the vicinity of the colloidal surface.
\bibitem{note_PRE_WC_curvature}
This is certainly justified if the counterion-counterion distance is smaller
  than the curvature radius of the macroion.
\end{thebibliography}

\begin{table}

\caption{Simulation parameters with some fixed values.}

\label{tab.simu-param}
\begin{tabular}{cc}
 parameters&
\\
\hline 
\( \sigma =3.57 \) \AA\ &
 Lennard Jones length units\\
 \( T_{0}=298K \)&
 room temperature\\
 \( \epsilon _{LJ}=k_{B}T_{0} \)&
 Lennard Jones energy units\\
 \( Z_{m} \)&
 macroion valence\\
 \( Z_{c}=2 \)&
 counterion valence\\
 \( l_{B}=10\sigma  \)&
 Bjerrum length\\
 \( f_{m} \)&
 macroion volume fraction\\
 \( a=8\sigma  \)&
 macroion-counterion distance of closest approach \\
\end{tabular}\end{table}

\begin{table}

\caption{Measured values for an \textit{isolated} macroion, with fixed radius \protect\( a\protect \),
of the energy gain for the first overcharging counterion \protect\( \Delta E^{OC}_{1}\protect \)
for various macroion bare charge \protect\( Z_{m}=2N_{c}\protect \). The value
of \protect\( \alpha \protect \) can be compared to the prediction of WC theory
for an infinite plane, which gives 1.96, compare text. We also record the values
of the fitting parameter \protect\( \tau \protect \) of Eq. (\ref{eq.Interaction-fit})
for selected \protect\( N_{c}\protect \) corresponding to those of Fig. (\ref{fig.tau-WC}).
The symbol {}``\protect\( ^{(\mathrm{i})}\protect \){}`` stands for the ionization
process discussed in Sec. \ref{sec.gs-2colAB}.}

\label{tab.WC}
\begin{tabular}{ccccc}
 \textit{\( Z_{m} \)}&
 \( N \)\( _{c} \)&
 \( \Delta E_{1}/k_{B}T_{0} \)&
\( \alpha  \)&
$\tau \sigma$\\
\hline 
2&
1&
-2.5&
1.94&
0.12\\
4&
2&
-3.8&
1.89&
0.18\\
6&
3&
-5.3&
1.97&
0.19\\
8&
4&
-6.1&
1.92&
0.24\\
10&
5&
-7.5&
2.02&
0.24\\
20&
10&
-10.7&
1.93&
-\\
30\( ^{(\mathrm{i})} \) &
 15&
+17.9 &
1.91 &
-\\
32&
16&
-&
-&
0.41\\
50&
25&
-18.0&
1.92&
0.51\\
 90&
 45&
-24.4 &
 1.88&
0.68\\
90\( ^{(\mathrm{i})} \)&
45&
+29.2&
1.89&
-\\
128&
64&
-&
-&
0.79\\
 150\( ^{(\mathrm{i})} \)&
 75&
+37.4 &
 1.91&
-\\
 180&
 90&
-35.3&
 1.88&
0.93\\
288&
144&
-&
-&
1.19\\
360&
180&
-50.0 &
1.86&
-\\
\end{tabular}\end{table}

\begin{table}

\caption{Measured values of the energy gain \protect\( \Delta E^{OC}_{1}\protect \)
and fixed counterion concentration \protect\( c\protect \), varying this time
the macroion radius \protect\( a\protect \) and the number of counterions \protect\( N_{c}\protect \). }

\label{tab.WC-curvature}
\begin{tabular}{cccc}
 $a/\sigma$&
 \( N \)\( _{c} \)&
 \( \Delta E_{1}/k_{B}T_{0} \)&
\( \alpha  \)\\
\hline 
6&
9&
-13.3&
1.91\\
8&
16&
-14.4&
1.97\\
10&
25&
-14.5 &
1.93\\
12&
36&
-14.7&
1.92\\
 14&
 49&
-15.1 &
 1.94\\
16&
64&
-15.1&
1.92\\
 20&
 100&
-15.3 &
 1.92\\
 40&
 400&
 -15.9&
1.94\\
80&
1600&
-16.4&
1.97\\
160&
6400&
-16.5&
1.96\\
\end{tabular}
\end{table}

\begin{table}

\caption{Measured value of the energy barrier and fit parameters of the electrostatic
interaction process involved in Fig. \ref{fig.barrier}(a) for \protect\( R/a=2.4\protect \)
and for for different macroion bare charges.}

\label{tab.fit-small-R}
\begin{tabular}{cccc}
 \textit{\( Z_{m} \)}&
 \( \Delta E_{bar}/k_{B}T_{0} \)&
 \( \Delta E^{*}_{1}/k_{B}T_{0} \)&
 \( \tau \sigma  \)\\
\hline 
50 &
 4.9&
 -20.4&
 0.51\\
 90&
 9.6&
 {\small -27.5}&
 0.68\\
 180&
 20.4&
 {\small -39.4}&
 0.92 \\
\end{tabular}\end{table}

\begin{table}

\caption{Measured value of the energy barrier and fit parameters of the electrostatic
interaction process involved in Fig. \ref{fig.barrier}(b) for \protect\( R/a=4.25\protect \)
and for different macroion bare charges.}

\label{tab.fit-large-R}
\begin{tabular}{cccc}
 \textit{\( Z_{m} \)}&
 \( \Delta E_{bar}/k_{B}T_{0} \)&
 \( \Delta E^{*}_{1}/k_{B}T_{0} \)&
 \( \tau \sigma  \)\\
\hline 
50 &
 16.8&
 -18.4&
 0.51\\
 90&
 23.3&
 {\small -24.4}&
 0.68\\
 180&
 33.8&
 {\small -35.3}&
 0.92 \\
\end{tabular}\end{table}

\begin{figure}
\caption{Ground state configurations for two, three, four and five counterions. The
corresponding geometrical figures show the typical angles. The electrostatic
energy (in units of \protect\( k_{B}Tl_{B}/a\protect \)) is given for a
central charge of +2\textit{e}. }
\label{fig.gillespie}
\end{figure}

\begin{figure}

\caption{Ground state surface counterion correlation functions for \protect\( Z_{m}=180\protect \)
and two states of charge {[}neutral (\protect\( n=0\protect \)) and overcharged
(\protect\( n=8\protect \)){]}. }

\label{fig.ground-state-CCF}
\end{figure}

\begin{figure}

\caption{Snapshot of the ground state structure of the neutral state (\protect\( n=0\protect \))
with a macroion charge \protect\( Z_{m}=180\protect \) {[}see Fig. \ref{fig.ground-state-CCF}
for the corresponding \protect\( g(r)\protect \){]}.}

\label{fig.gs-snapshot-OC_0}
\end{figure}

\begin{figure}

\caption{Surface counterion correlation functions at \textit{room temperature} \protect\( T_{0}\protect \)
for two states of charge {[}neutral (n=0) and overcharged (n=8){]} with \textit{\protect\( Z_{m}=180\protect \)}
and \protect\( f_{m}=6.6\times 10^{-3}\protect \). }

\label{fig.temperature-CCF}
\end{figure}

\begin{figure}

\caption{Electrostatic energy (in units of \protect\( k_{B}T_{0}\protect \)) for \textit{ground
state} configurations of a single charged macroion of as a function of the number
of \textit{overcharging} counterions \protect\( n\protect \) for three different
bare charges \textit{\protect\( Z_{m}\protect \)}. The neutral case was chosen
as the potential energy origin, and the curves were produced using the theory
of Eq. (\ref{Eq.WC-n-OC}), compare text.}

\label{fig.OC-MD-energy}
\end{figure}

\begin{figure}

\caption{Wigner crystal parameter \protect\( \alpha \protect \) as a function of the
number of counterions \protect\( N_{c}\protect \) for fixed colloid radius
\protect\( a\protect \).}

\label{fig.alpha-WC-N}
\end{figure}

\begin{figure}

\caption{Wigner crystal parameter \protect\( \alpha \protect \) as a function of the
colloid radius \protect\( a\protect \) for a fixed surface counterion concentration
\protect\( c\protect \).}

\label{fig.alpha-WC-a}
\end{figure}

\begin{figure}

\caption{Electrostatic interaction energy (in units of \protect\( k_{B}T_{0}\protect \))
of a divalent counterion with a neutral effective colloid {[}bare particle +
surrounded counterions{]} as function of distance \protect\( r/a\protect \)
from the center of a macroion for two different macroion bare charges \protect\( Z_{m}\protect \).
The energy is set to zero at distance infinity. Solid lines correspond to exponential
fits {[}see Eq. (\ref{eq.Interaction-fit}){]}.}

\label{fig.gs-interaction-PROFILE}
\end{figure}

\begin{figure}

\caption{Exponential fit parameter \protect\( \tau \protect \) as a function of the
square root of the number of counterions \protect\( \sqrt{N_{c}}\protect \).
The dashed line corresponds to a linear fit in \protect\( \sqrt{N_{c}}\protect \).}

\label{fig.tau-WC}
\end{figure}

\begin{figure}

\caption{Snapshot of a {}``pseudo-equilibrium{}'' configuration at room temperature
\protect\( T_{0}\protect \) where the counterion-layers do not exactly compensate
the macroions charge. Here the deficiency charge is \protect\( \pm 1\protect \)
counterion (or \protect\( \pm 2e\protect \) as indicated above the macroions)
and \textit{R}/\textit{a} = 3.6\textit{.}}

\label{fig.snapshot-179-181}
\end{figure}

\begin{figure}

\caption{Total electrostatic energy (in units of \protect\( k_{B}T_{0}\protect \))
of the system, for \textit{zero temperature} configurations, of two macroions
at a center-center separation of (a) \protect\( R/a=2.4\protect \) (b) \protect\( R/a=4.25\protect \)
as a function of one displaced counterion distance from the left macroion for
three typical values \protect\( Z_{m}\protect \). The exact neutral state was
chosen as the potential energy origin. The schematic drawing indicates the path
(dotted line) of the moved counterion. The ending arrows of the arc indicate
the start position (left sphere) and final position (right sphere) of the moved
counterion. Dashed lines correspond to the fit using Eq. (\ref{eq.barrier-fit-2col})
of which parameters can be found in Tables \ref{tab.fit-small-R} and \ref{tab.fit-large-R}. }

\label{fig.barrier}
\end{figure}

\begin{figure}

\caption{Projection of the counterion positions, located on both inner (face to face)
hemispheres, along the symmetrical axis passing through the macroion centers.
Open (filled) circles are counterions belonging to macroion A (B). One clearly
sees the interlocking of the two ordered structures yielding locally to a superlattice. }

\label{fig.superlattice}
\end{figure}

\begin{figure}

\caption{Reduced effective force between the two spherical macroions at \textit{zero
temperature} for \protect\( Z_{m}=180\protect \) as a function of distance
from the center. The different forces are explained in the text. The lines are
a guide to the eye.}

\label{fig.Effective-Force}
\end{figure}

\begin{figure}

\caption{Total electrostatic energy as a function of the degree of ionization for zero
temperature configurations of two colloids (\protect\( A\protect \) and \protect\( B\protect \)),
for three typical charges \protect\( Z_{B}\protect \) (30, 90 and 150) for
macroion \textit{B} and for three given distance separations: (a) \protect\( R/a=4.25\protect \),
(b) \protect\( R/a=3.0\protect \) and (c) \protect\( R/a=2.4\protect \). Dashed
lines were obtained using Eq. (\ref{eq.finiteR}).}

\label{fig.gs-DI}
\end{figure}

\begin{figure}

\caption{Total electrostatic energy as a function of the degree of ionization for zero
temperature configurations of an \textit{isolated} colloid. The three upper
curves correspond to the ionization energy for the three typical charges \protect\( Z_{B}\protect \)
(30, 90 and 150). The lower curve corresponds to the energy gained (changed
sign for commodity) by overcharging \protect\( (Z_{A}=180)\protect \). Dashed
lines were obtained using Eqs. (\ref{Eq.WC-n-OC}, \ref{Eq.WC-Ionization})
with the measured values for \protect\( \alpha \protect \) from Table \ref{tab.WC}.}

\label{fig.Ionization}
\end{figure}

\begin{figure}

\caption{Relaxation, at room temperature \protect\( T_{0}=298K\protect \), of an initial
unstable neutral state towards ionized state. Plotted is the total electrostatic
energy versus time (LJ units), for \protect\( Z_{B}=30\protect \) and \protect\( R/a=2.4\protect \).
Dashed lines lines represent the mean energy for each \textit{DI} state. Each
jump in energy corresponds to a counterion transfer from the macroion \textit{B}
to macroion \textit{A} leading to an ionized state (\protect\( DI=2\protect \))
which is lower in energy than the neutral one. The two energy jumps 
\protect\( \Delta E_{1}/k_{B}T_{0}=-19.5\protect \)
and \protect\( \Delta E_{2}/k_{B}T_{0}=-17.4\protect \) are in very good agreement
with those of Fig. \ref{fig.gs-DI}(c) (-20.1 and -16.3).}

\label{fig.relaxation}
\end{figure}

\begin{figure}

\caption{Snapshot of the ionized state (\protect\( DI=2\protect \)) obtained in the
relaxation process depicted in Fig. \ref{fig.relaxation}, with the net charges
\textit{\protect\( +4e\protect \)} and \textit{\protect\( -4e\protect \)}
as indicated.}

\label{fig.snaphot-DI2}
\end{figure}

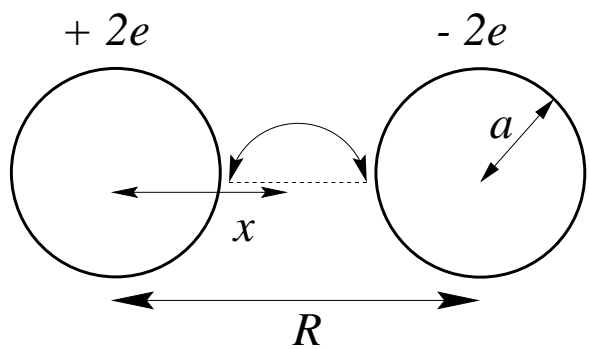
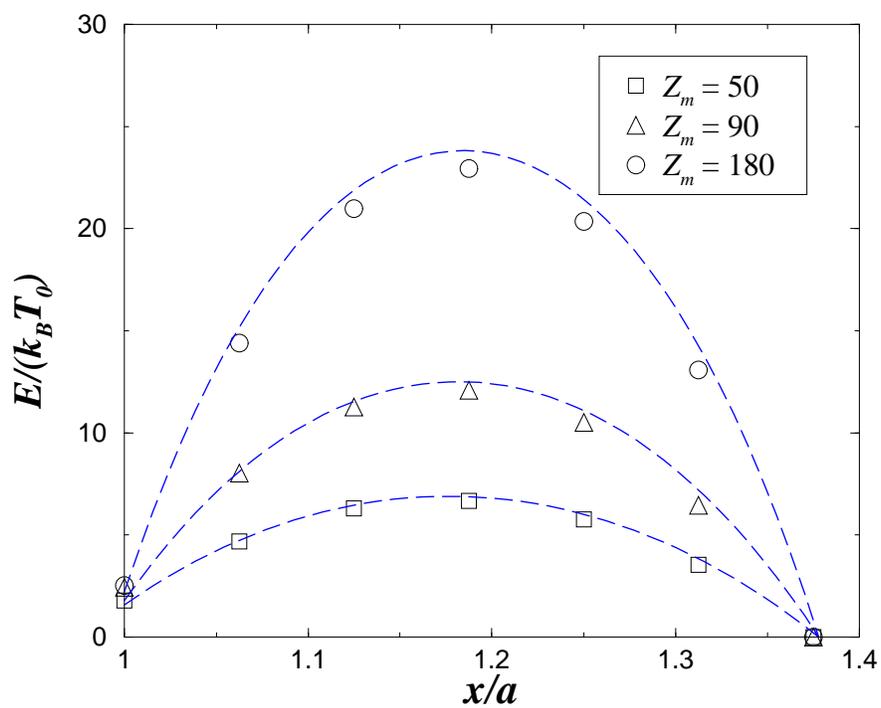
\end{document}